\documentclass[twocolumn,showpacs,preprintnumbers,amsmath,amssymb, superscriptaddress]{revtex4-1}

\usepackage{natbib}
\usepackage{graphicx}
\usepackage{dcolumn}
\usepackage{dsfont}
\usepackage{bm}
\usepackage{float}
\newtheorem{theorem}{Theorem}

\begin{document}

\title{Universal simulation of Markovian open quantum systems}

\date{\today}

\author{Ryan Sweke}
\email{rsweke@gmail.com}
\affiliation{Quantum Research Group, School of Physics and Chemistry, University of KwaZulu-Natal, Durban, 4001, South Africa.}
\author{Ilya Sinayskiy}
\affiliation{Quantum Research Group, School of Physics and Chemistry, University of KwaZulu-Natal, Durban, 4001, South Africa.}
\affiliation{National Institute for Theoretical Physics (NITheP), KwaZulu-Natal, South Africa.}
\author{Denis Bernard}
\affiliation{Laboratoire de Physique Th\'{e}orique de l'ENS, CNRS and Ecole Normale Sup\'{e}rieure de Paris, France.}
\author{Francesco Petruccione}
\affiliation{Quantum Research Group, School of Physics and Chemistry, University of KwaZulu-Natal, Durban, 4001, South Africa.}
\affiliation{National Institute for Theoretical Physics (NITheP), KwaZulu-Natal, South Africa.}

\begin{abstract}
We consider the problem of constructing a ``universal set'' of Markovian processes, such that any Markovian open quantum system, described by a one-parameter semigroup of quantum channels, can be simulated through sequential simulations of processes from the universal set. In particular, for quantum systems of dimension $d$, we explicitly construct a universal set of semigroup generators, parametrized by $d^2-3$ continuous parameters, and prove that a necessary and sufficient condition for the dynamical simulation of a $d$ dimensional Markovian quantum system is the ability to implement a) quantum channels from the semigroups generated by elements of the universal set of generators, and b) unitary operations on the system. Furthermore, we provide an explicit algorithm for simulating the dynamics of a Markovian open quantum system using this universal set
of generators, and show that it is efficient, with respect to this universal set, when the number of distinct Lindblad operators (representing physical dissipation processes) scales polynomially with respect to the number of
subsystems.

\end{abstract}

\pacs{03.67.Ac, 03.65.Yz, 89.70.Eg}
\maketitle

\section{Introduction}\label{intro}

All quantum systems are invariably in contact with some environment to some extent. As a result, the development of tools for the study of such open quantum systems, undergoing non-unitary dynamics as a result of system-environment interactions, is of importance for understanding a rich variety of phenomena \cite{Petruccione2002, Daley2014}. In particular, the study of open quantum systems allows us to better understand the nature of dissipation and decoherence \cite{Petruccione2002, Daley2014}, thermalisation and equilibration \cite{Znidaric2010, Kastoryano2014}, non-equilibrium phase transitions \cite{Pizorn2013, Prosen2009} and transport phenomena in both strongly-correlated \cite{Prosen2011, Benenti2009, Prosen2012} and biological systems \cite{Marais2013, Marais2012, Dorner2012}. Furthermore, it has been shown that dissipation and decoherence, traditional enemies of quantum information processing, can be exploited as a resource for quantum computation \cite{Verstraete2009, Sinayskiy2012}, the preparation of topological phases \cite{Budich2014, Diehl2011, Bardyn2013} and the preparation of entangled states \cite{Kraus2008, Diehl2008}.

Simulations on controllable quantum devices promise to be one of the most effective tools for the study of open quantum systems, and while the majority of effort over the past twenty years has focused on the development of methods for the simulation of closed quantum systems \cite{Sanders2013, Brown2010, Georgescu2014, Berry2015}, which undergo Hamiltonian generated unitary evolution, a plethora of methods have also been developed for the quantum simulation of open quantum systems, on a wide variety of quantum devices. These methods include  collision model based approaches \cite{Hillery2002, Koniorczyk2006, Koniorczyk2005, Ziman2011, Rybar2012}, simulation algorithms designed for conventional unitary gate based universal quantum computers \cite{Lloyd1996, Terhal1999, Terhal2000, Bacon2001, Temme2011, Poulin2009, Chiang2010, Yung2012, Kliesch2011, Barthel2012, Tempel2014, DiCandia2014, Wang2013, Wang2014, Wang2011, Sweke2014} and simulation algorithms designed for more general quantum simulators incorporating feedback and dissipative elements in addition to unitary gates \cite{Lloyd2001, Muller2011, Schindler2013, Barreiro2011, Weimer2011, Weimer2010, Herdman2010, Young2012}.

However, despite the wide variety of methods for the simulation of open quantum systems, there exists no ``universal set" of non-unitary processes through which all such processes can be simulated via sequential simulations from the universal set. This is in clear contrast with the situation for Hamiltonian generated unitary evolution, for which it is well known that any unitary operation can be implemented, up to arbitrary precision, using some (not necessarily efficient) sequence of unitary gates from a finite universal set \cite{Nielsen2000}. Such universal sets are interesting not only from a fundamental perspective, but also from a pragmatic perspective, as they allow for experimental development to be focused on developing the capability of implementing a reduced set of significantly simpler processes.

One natural response to this problem is via the Stinespring dilation \cite{Stinespring1955}. Given any non-unitary dynamics of some particular system, it is always possible to introduce some environment, with size the square of the system size in the general case, such that the non-unitary dynamics of the system may be simulated through unitary evolution  of the total system and environment \cite{Lloyd1996, Terhal1999, Kliesch2011, Wang2013, Wang2014, Sweke2014}. However,  it is important to note that for an arbitrary non-unitary process, there is no guarantee that the dilated unitary admits an \textit{efficient} decomposition into some sequence of unitary gates from a universal set \cite{Nielsen2000}, and as such this strategy offers an advantage for the construction of efficient simulation algorithms only when the original non-unitary process exhibits some useful structure, such as local interactions \cite{Kliesch2011}. Furthermore, in line with the spirit of dissipative state preparation \cite{Kraus2008, Diehl2008}, we would like to investigate the possibility of developing a universal set which might allow us to exploit the natural dissipation and decoherence present in any controlled quantum device.

Therefore, as an alternative approach,  one can consider the problem of identifying the smallest set of non-unitary dynamics, applied to the system only, such that if one has the resources to simulate dynamics from this set, and implement unitary operations on the system, then one will be able to simulate any non-unitary dynamics up to arbitrary precision. This problem has been considered before. In particular, Wang et al. have constructed a method for the simulation of arbitrary quantum channels through the simulation of extreme channels \cite{Wang2013, Wang2014}, and in effect identified such a universal set for discrete time evolution of open quantum systems. However, for systems evolving continuously in time, even in the simplest case of Markovian semigroup dynamics it is necessary to first exponentiate the generator of the semigroup in order to obtain the quantum channels describing time evolution. This is infeasible for an arbitrary semigroup generator and in order to address this problem Bacon et al. \cite{Bacon2001} have constructed a composition framework for the combination and transformation of semigroup generators. Using this framework they were able to identify a continuous one-parameter set of semigroup generators and demonstrate that one can efficiently simulate arbitrary Markovian dynamics of a single qubit through simulations of quantum channels from the semigroups generated by this one parameter set of generators \cite{Bacon2001, Sweke2014}.

Despite this initial progress, extending these results to arbitrary Markovian open quantum systems has remained a challenging open problem. In this work we address this problem by using the composition framework of \cite{Bacon2001} to construct a continuous $d^2-3$ parameter set of generators, which is universal in the sense that given the ability to implement quantum channels from the semigroups generated by elements of this set of generators, along with unitary operations on the system, one can simulate the dynamics of an arbitrary $d$ dimensional Markovian quantum system up to arbitrary precision. This set of generators is minimal within this particular composition framework, and by construction of this set we complete the program initiated in \cite{Bacon2001}, proving that the dimension of this universal set is indeed as originally conjectured.

Furthermore, assuming the ability to implement unitary operations on the system along with quantum channels from the semigroups generated by elements of the universal set, we utilise recent error bounds for superoperator Suzuki-Lie-Trotter expansions \cite{Sweke2014} to construct an explicit algorithm for the simulation of arbitrary Markovian open quantum systems, and analyse the conditions under which a Markovian open quantum system may be efficiently simulated, with respect to the constructed universal set, using this algorithm. 

This paper is structured as follows: We begin in Section \ref{setting} by introducing the formalism of Markovian semigroup dynamics and formulating the problem of simulating such dynamics. We then proceed, in Section \ref{compsec}, to introduce the composition framework of \textit{linear combination} and \textit{unitary conjugation}, introduced in \cite{Bacon2001}, for the combination of Markovian semigroup generators. Given this framework, we then present our main result in Section \ref{mainresult}, the construction of a universal set of generators for arbitrary Markovian dynamics. A detailed proof of the main result is then given in Section \ref{proof}, before discussing in Section \ref{simsec} the consequences for simulation of Markovian open quantum systems.

\section{Setting}\label{setting}

Given a quantum system with Hilbert space $\mathcal{H}_S \cong \mathbb{C}^d$, we are concerned with Markovian semigroup dynamics, in which the state of the system $\rho(t)\in\mathcal{B}(\mathcal{H}_S)$ evolves according to a quantum Markov master equation

\begin{equation}\label{master}
\frac{d}{dt}\rho(t) = \mathcal{L}\rho(t),
\end{equation}
where $\mathcal{L} \in \mathcal{B}(\mathcal{B}(\mathcal{H}_S))$ is the generator of a uniformly continuous one parameter semigroup of quantum channels $\{T(t)\}$, which we refer to as a Markovian semigroup \cite{Petruccione2002}. The state of the system at time $t>t_0$ is then given by $\rho(t) = T(t-t_0)\rho(t_0) = e^{(t-t_0)\mathcal{L}}\rho(t_0)$. Furthermore, \eqref{master} may always be written in the form

\begin{equation}\label{GKS}
\mathcal{L}(\rho) =i[\rho,H] + \sum_{l,k = 1}^{d^2-1}A_{l,k}\bigg(F_l\rho F_k^{\dagger} - \frac{1}{2}\lbrace F_k^{\dagger}F_l,\rho   \rbrace_+   \bigg),
\end{equation}
for some Hermitian operator $H = H^\dagger \in \mathcal{M}_d(\mathbb{C})$ and some positive semidefinite  $A \in \mathcal{M}_{d^2-1}(\mathbb{C}) $, where $\{F_i\}$ is some basis for the space of traceless matrices in $\mathcal{M}_d(\mathbb{C})$, and without loss of generality from this point we will always utilise the Hermitian traceless basis which generalises the Gell-Mann basis for $\mathrm{su}(3)$. Eq. \eqref{GKS} is known as the Gorini, Kossakowski, Sudarshan and Lindblad (GKSL) form  of the quantum Markov master equation and we refer to $A$ as the GKS matrix. Additionally, note that via diagonalisation of the GKS matrix $A$, Eq. \eqref{GKS} can always be brought into, and is often specified in, the so called diagonal form,

\begin{equation}\label{diagonalform}
\mathcal{L}(\rho) =i[\rho,H] + \sum_{k = 1}^{m}\gamma_k \bigg(L_k \rho L_k^{\dagger} - \frac{1}{2}\lbrace L_k^{\dagger}L_k,\rho   \rbrace   \bigg),
\end{equation}
where $m$ is the number of non-zero eigenvalues of $A$, and typically each Lindblad operator $L_k$ represents some physical dissipation process \cite{Petruccione2002}.

In order to discuss simulations of Markovian semigroups it is necessary to have some means for quantifying the error in approximations of generators and quantum channels. To achieve this we will utilise the $(1\rightarrow 1)$-norm for super-operators, where in general the $(p\rightarrow q)$-norm of a super-operator $T \in \mathcal{B}(\mathcal{B}(\mathcal{H}))$ is defined as \cite{Watrous} 

\begin{equation}
||T||_{p\rightarrow q} := \sup_{||A||_p=1}||T(A)||_q.
\end{equation}
The $(p\rightarrow q)$-norm defined above is induced from the Schatten $p$-norm of an operator, defined as $||A||_p:= \big(\mathrm{tr}(|A|^p)\big)^{\frac{1}{p}}$ for all $A \in \mathcal{B}(\mathcal{H})$. We use the $(1\rightarrow 1)$-norm as this is induced by the Schatten 1-norm, which corresponds up to a factor of 1/2 with the trace distance, $\mathrm{dist}(\rho,\sigma):=\sup_{0\leq A\leq 1}\mathrm{tr}\big(A(\rho-\sigma)\big)$, arising from a physical motivation of operational distinguishability of quantum states \cite{Nielsen2000}, which is relevant when working in the Schr\"{o}dinger picture.

At this stage, given a Markovian semigroup $\{T(t)\}$, generated by $\mathcal{L} \in \mathcal{B}(\mathcal{B}(\mathcal{H}_S))$ with $\mathrm{dim}(\mathcal{H}_s) = d$, we say that the semigroup can be \textit{efficiently simulated} if given any initial state $\rho(0) \in \mathcal{B}(\mathcal{H_S})$, any $\epsilon > 0$ and any $t>0$, there exists a well defined procedure, requiring at most $\mathrm{poly}\big(||\mathcal{L}||_{(1\rightarrow 1)},t,1/\epsilon, \mathrm{ln}(d)\big)$ applications of standard resources, such that the output of the procedure is a state $\tilde{\rho}$ satisfying $||\tilde{\rho} - \rho(t)||_1 < \epsilon$. Note that $\mathrm{poly}$ denotes any polynomial function and that for many-body systems $\mathrm{ln}(d)$ is proportional to the number of subsystems. Furthermore, note that the standard resources depend on the simulator on which the well defined procedure, or algorithm, is executed. If we are considering simulations on a universal quantum computer, then the procedure would be a quantum circuit, and the resources would be unitary gates from some finite universal set. However, motivated by the spirit of dissipative state preparation, in this paper we are considering more general simulators whose standard resources might include additional non-unitary elements capable of exploiting natural or engineered dissipation. In particular, under the understanding that we are considering this more general context, we will consider as standard resources all quantum channels from semigroups generated by elements of the universal set constructed in Section \ref{mainresult}, in addition to arbitrary unitary operations.

\section{Composition Framework}\label{compsec}

In this section, following \cite{Bacon2001}, we present a composition and transformation framework through which one can combine and transform the generators of Markovian semigroups to form the generator of a new Markovian semigroup.  As described in Section \ref{intro}, this composition framework will allow us to identify in Section \ref{mainresult} a parametrized universal set of semigroup generators, through which all Markovian semigroups of a given dimension can be simulated, up to arbitrary precision. 

This composition framework consists of two procedures, \textit{linear combination} and \textit{unitary conjugation}. Firstly, let $\mathcal{L}_a$ and $\mathcal{L}_b$ be the generators of Markovian semigroups $\{T^{(a)}(t)\}$ and $\{T^{(b)}(t)\}$ respectively. The \textit{linear combination} of $\mathcal{L}_a$ and $\mathcal{L}_b$ is then quite simply defined as the super-operator $\mathcal{L}_{a+b} = \mathcal{L}_a + \mathcal{L}_b$, the generator of a Markovian semigroup $\{T^{(a+b)}(t)\}$ \cite{Bacon2001}. From a generalisation of the Lie-Trotter theorem \cite{Trotter1959} into the superoperator regime \cite{Sweke2014, Kliesch2011}, we see that

\begin{equation}\label{lielimit}
T^{(a+b)}(t) =e ^{t\mathcal{L}_{a+b}} =  \lim_{n \rightarrow \infty} \big[T^{(a)}(t/n)T^{(b)}(t/n) \big]^n.
\end{equation}
The generalisation of this procedure to the linear combination of multiple generators is then straightforward. Furthermore, as discussed in detail in Appendix \ref{lincomb}, using Suzuki-Lie-Trotter techniques  \cite{Suzuki1990, Suzuki1991}, generalised from the context of Hamiltonian simulation \cite{Papageorgiou2012, Wiebe2010}, one can show that the infinite sum in Eq. \eqref{lielimit} can be effectively truncated, such that any channel from the semigroup generated by the linear combination $\mathcal{L}_{a+b}$ can be implemented, up to arbitrary precision, through a finite number of implementations of channels from the semigroups generated by the constituent generators
$\mathcal{L}_a$ and $\mathcal{L}_b$ \cite{Sweke2014, Kliesch2011}. A discussion of when the Markovian semigroup generated by the linear combination of multiple generators can be \textit{efficiently} simulated is postponed until Section \ref{simsec}.

Note that given any generator $\mathcal{L}$ we can always rewrite \eqref{GKS} as

\begin{equation}
\mathcal{L}(\rho) = \mathcal{L}_H(\rho) + \mathcal{L}_A(\rho),
\end{equation}
where
\begin{equation}
\mathcal{L}_H(\rho) = i[\rho,H] 
\end{equation}
and
\begin{equation}
\mathcal{L}_A(\rho) = \sum_{l,k = 1}^{d^2-1}A_{l,k}\bigg(F_l\rho F_k^{\dagger} - \frac{1}{2}\lbrace F_k^{\dagger}F_l,\rho   \rbrace_+   \bigg).
\end{equation}
Therefore, if we assume the ability to implement arbitrary unitary operations on the system, then without loss of generality we can set $H = 0$, as we can always reintroduce the unitary contribution and implement the total generator $\mathcal{L}$ through linear combination of $\mathcal{L}_H$ and $\mathcal{L}_A$.

The second transformation procedure, \textit{unitary conjugation}, is defined as follows: Given a Hilbert space $\mathcal{H}_S \cong \mathbb{C}^d $  and a Markovian semigroup $\{T(t)\}$ with generator $\mathcal{L} \in \mathcal{B}(\mathcal{B}(\mathcal{H}_S))$, for any unitary operator $U \in \mathrm{SU}(d)$ the unitary conjugation via $U$ of the semigroup $\{T(t)\}$ is the new Markovian semigroup 

\begin{equation}
\{T_{U}(t)\} \equiv \{\mathcal{U}^{\dagger}T(t)\mathcal{U}\},
\end{equation}
where $\mathcal{U}(\rho) = U\rho U^{\dagger}$. The following theorem, due to \cite{Bacon2001}, is particularly important, as it describes the manner in which the GKS matrix specifying $\mathcal{L}$ is transformed as a result of unitary conjugation of the semigroup $\{T(t)\}$. The statement of this theorem relies on notions related to the adjoint representation of a Lie group, presented in detail in Appendix \ref{adjointap}. Note in particular that $\mathrm{Int}\big(\mathrm{su}(d)\big)$ denotes the image of the adjoint representation of $\mathrm{SU}(d)$, a Lie group itself, while $\mathfrak{Int}\big(\mathrm{su}(d)\big)$ is the Lie algebra of $\mathrm{Int}\big(\mathrm{su}(d)\big)$. 

\begin{theorem}\label{unconj}
Assume $\mathcal{H}_S \simeq \mathbb{C}^d$ and that $\mathcal{L} \in \mathcal{B}(\mathcal{B}(\mathcal{H}_S))$ is the generator of a Markovian semigroup with $H = 0$, such that

\begin{align}
\mathcal{L}(\rho) &= \mathcal{L}_A(\rho)\\
&=\sum_{l,k = 1}^{d^2-1}A_{l,k}\bigg(F_l\rho F_k^{\dagger} - \frac{1}{2}\lbrace F_k^{\dagger}F_l,\rho   \rbrace_+   \bigg).
\end{align}
Furthermore, assume that $\{F_\gamma\}|_{\gamma = 1}^{d^2-1}$ is a Hermitian basis for the space of traceless matrices in $\mathcal{M}_d(\mathbb{C})$, such that $\{iF_\gamma\}|_{\gamma = 1}^{d^2-1}$ is a basis for $\mathrm{su}(d)$ and $U = \mathrm{exp}\big(\sum_{\gamma = 1}^{d^2-1}ir_\gamma F_{\gamma}\big) \in \mathrm{SU}(d)$ for any $\vec{r} \in \mathbb{R}^{d^2-1}$. Then,

\begin{align}
\mathcal{U}^{\dagger}T_t\mathcal{U} &= \mathcal{U}^{\dagger}e^{t\mathcal{L}_A}\mathcal{U} \\
&= e^{t \mathcal{L}_{\tilde{A}}}\\
&= T_{U}(t),
\end{align}
where,
\begin{equation}
\mathcal{L}_{\tilde{A}}(\rho) = \sum_{l,k = 1}^{d^2-1}\tilde{A}_{l,k}\bigg(F_l\rho F_k^{\dagger} - \frac{1}{2}\lbrace F_k^{\dagger}F_l,\rho   \rbrace_+   \bigg),
\end{equation}
with $\tilde{A} = G_{(U)}AG_{(U)}^T$, where $G_{(U)} \in \mathrm{Int}\big(\mathrm{SU}(d)\big)$ is given by

\begin{equation}
G_{(U)} = \hat{\mathrm{Ad}}\big(U\big) = \mathrm{exp}\big(\sum_{\gamma =1 }^{d^2-1} i r_\gamma G_\gamma\big), 
\end{equation}
and $\{iG_{\gamma}\}$ is a basis for $\mathfrak{Int}\big(\mathrm{su}(d)\big)$, with matrix elements $[G_\gamma]_{\alpha\beta} = if_{\gamma\alpha\beta}$, where $f_{\gamma\alpha\beta}$ are the real structure constants of $\mathrm{su}(d)$, defined via 
\begin{equation}
[F_{\gamma},F_{\alpha}] = i\sum_{\beta = 1}^{d^2-1}f_{\gamma\alpha\beta}F_{\beta}. 
\end{equation}

\end{theorem}

Colloquially, Theorem \ref{unconj} states that unitary conjugation of the semigroup results in conjugation of the GKS matrix by an element of the adjoint representation of $\mathrm{SU}(d)$. As such, we see that by adding together the generators of Markovian semigroups (linear combination), or conjugating the generators via elements of $\mathrm{Int}(\mathrm{SU}(d))$ (unitary conjugation), we obtain the generators of new Markovian semigroups which can be simulated (though perhaps not necessarily efficiently), provided the semigroups corresponding to the original constituent generators can be simulated and arbitrary unitary operations can be implemented on the system.

\section{Main Result}\label{mainresult}

Given the composition framework of Section \ref{compsec}, we can now present our main result, the construction of a universal set of Markovian semigroup generators, parameterised by $d^2-3$ continuous parameters, for Markovian open quantum systems of any dimension $d$. For $d = 2$ this set was first constructed in \cite{Bacon2001}, and our construction, presented as Theorem \ref{main}, generalises this original method to arbitrary dimension. As per the statement of the theorem, the constructed set is universal with respect to the composition framework of linear combination and unitary conjugation, i.e. universal in the sense that in order to simulate any Markovian semigroup it is necessary and sufficient to be able to implement arbitrary unitary operations on the system, along with all quantum channels from the semigroups generated by the $d^2-3$ parameter family of generators. It is important to note however that, as in the unitary case, if we consider operations from the universal set as our ``standard resources", we do not necessarily expect to be able to \textit{efficiently} simulate all Markovian semigroups in terms of these resources. In Section \ref{simsec} we utilise the construction of the proof of Theorem \ref{main}, presented in Section \ref{proof}, to construct an explicit algorithm for the (not necessarily efficient) simulation of an arbitrary Markovian semigroup via simulations of semigroups from the universal set, and then analyse the conditions under which a class of Markovian open quantum systems may be efficiently simulated using this particular algorithm.

\begin{theorem}\label{main}
In order to simulate, using linear combination and conjugation by unitaries, an arbitrary Markovian semigroup generated by $\mathcal{L} \in \mathcal{B}(\mathcal{B}(\mathcal{H}_S))$ with $\mathcal{H}_{S}\simeq \mathbb{C}^d$, it is necessary and sufficient to be able to simulate all Markovian semigroups whose generator is specified by a GKS matrix from the $d^2 - 3$ parameter family

\begin{equation}\label{family}
A(\theta,\vec{\alpha}^R,\vec{\alpha}^{I}) = \vec{a}(\theta,\vec{\alpha}^R,\vec{\alpha}^{I})\vec{a}(\theta,\vec{\alpha}^R,\vec{\alpha}^{I})^{\dagger},
\end{equation}
where
\begin{equation}\label{veccon}
\vec{a}(\theta,\vec{\alpha}^R,\vec{\alpha}^{I}) = \cos(\theta)\tilde{a}^R(\vec{\alpha}^R) + i\sin(\theta)\tilde{a}^I(\vec{\alpha}^{I})
\end{equation}
for $\theta \in [0,\pi/4]$, with $\tilde{a}^R(\vec{\alpha}^R), \tilde{a}^I(\vec{\alpha}^I) \in \mathbb{R}^{d^2-1}$ given by
\begin{equation}\label{firstform}
\tilde{a}^R(\vec{\alpha}^R) = \begin{pmatrix}
a^R_1 \\
\vdots \\
a^R_{d-1}\\
0\\
\vdots\\
\vdots\\
0
\end{pmatrix} \qquad \tilde{a}^I(\vec{\alpha}^I) = \begin{pmatrix}
a^I_1 \\
\vdots \\
\vdots\\
a^I_{d^2-d}\\
0\\
\vdots\\
0
\end{pmatrix},
\end{equation}
with
\begin{align}
&|\tilde{a}^R(\vec{\alpha}^R)| = |\tilde{a}^I(\vec{\alpha}^I)| =1 \label{norm}\\
&\tilde{a}^R(\vec{\alpha}^R)\cdot\tilde{a}^I(\vec{\alpha}^I) = 0, \label{orthog}
\end{align}
such that for $d \geq 3$,
\begin{align}
&a^R_1 = \cos(\alpha^R_1) \label{firstone}\\
&a^R_2 = \sin(\alpha^R_1)\cos(\alpha^R_2)\\
& \quad\vdots\nonumber\\
&a^R_{d-2} = \sin(\alpha^R_1) \ldots \sin(\alpha^R_{d-3})\cos(\alpha^R_{d-2})\\
&a^R_{d-1} = \sin(\alpha^R_1) \ldots \sin(\alpha^R_{d-3})\sin(\alpha^R_{d-2})
\end{align}
and
\begin{align}
&a^I_1 = \cos(\alpha^I_1) \\
&a^I_2 = \sin(\alpha^I_1)\cos(\alpha^I_2)\\
& \quad\vdots\nonumber\\
&a^I_{d^2 - d -1} = \sin(\alpha^I_1) \ldots \sin(\alpha^I_{d^2-d-2})\cos(\alpha^I_{d^2-d-1})\\
&a^I_{d^2 - d} = \sin(\alpha^I_1) \ldots \sin(\alpha^I_{d^2-d-2})\sin(\alpha^I_{d^2-d-1})
\end{align}
where, 
\begin{align}
&\alpha^R_j \in [0,\pi] \quad \mathrm{for}\quad j \in [1,d-3], \\
&\alpha^I_k \in [0,\pi] \quad \mathrm{for}\quad k \in [1,d^2-d- 2],\\
&\alpha^R_{d-2} \in [0,2\pi], \\
&\alpha^I_{d^2-d-1} \in [0,2\pi],
\end{align}
and

\begin{equation}\label{lastone}
\cos(\alpha^I_1) = \frac{1}{a^R_1}\bigg(\sum_{j =2}^{d-1} a^R_j a^I_j  \bigg)
\end{equation}
is constrained by orthogonality, and for $d = 2$,

\begin{equation}
\tilde{a}^R(\vec{\alpha}^R) = \begin{pmatrix}
1\\
0 \\
0
\end{pmatrix} \qquad \tilde{a}^I(\vec{\alpha}^I) = \begin{pmatrix}
0 \\
1 \\
0
\end{pmatrix}.
\end{equation}
\end{theorem}

\section{Proof of Theorem 2}\label{proof}
\subsection{Proof of sufficiency} Firstly, without any loss of generality we assume $H = 0$. Let $A \geq 0 \in \mathcal{M}_{d^2-1}(\mathbb{C})$ then be the GKS matrix specifying the generator of the Markovian semigroup we wish to simulate. $A$ is positive semidefinite and therefore via the spectral decomposition one can express $A$ as

\begin{equation}
A = \sum_{k}^m\lambda_k\vec{a}_{k}\vec{a}^{\dagger}_{k},
\end{equation}
where $\lambda_{k} \geq 0$, $m$ is the number of non-zero eigenvalues of $A$ and $|\vec{a}_{k}| = 1$ for all $k$. By linear combination it is therefore sufficient to be able to simulate all GKS matrices $\vec{a}\vec{a}^{\dagger}$ with $|\vec{a}| = 1$. Any such vector $\vec{a}$ can be split into real and imaginary part,

\begin{equation}
\vec{a} = \vec{a}^{R} + i\vec{a}^{I},
\end{equation}
where $\vec{a}^R, \vec{a}^I \in \mathbb{R}^{d^2-1}$. Furthermore, $\vec{a}$ appears only in outer products and as such the phase of $\vec{a}$ is irrelevant, i.e. if we define $\vec{a}' = e^{i\psi}\vec{a}$, then we see that  $\vec{a}\vec{a}^{\dagger} = \vec{a}'\vec{a}'^{\dagger}$, and therefore to simulate $\vec{a}\vec{a}^{\dagger}$ we could simulate $\vec{a}'\vec{a}'^{\dagger}$ for any value of $\psi$.  If we now define the two parameters

\begin{align}
k_1 &\equiv |\vec{a}^{R}|^2 - |\vec{a}^{I}|^2 \label{k1} \\
k_2 &\equiv 2 {\vec{a}^{R}} \cdot\vec{a}^I, \label{k2}
\end{align}
then we can see that a phase transformation
\begin{align}
\vec{a}' &= e^{i\psi}\vec{a} \\&= (\vec{a}^R\cos{\psi} - \vec{a}^{I}\sin{\psi}) + i(\vec{a}^R\sin{\psi} + \vec{a}^{I}\cos{\psi})
\end{align}
maps $k_1$ and $k_2$ according to
\begin{equation}\label{transformation}
\begin{pmatrix}
k_1' \\
k_2'
\end{pmatrix} = 
\begin{pmatrix}
\cos{2\psi} & -\sin{2\psi} \\
\sin{2\psi} & \cos{2\psi}
\end{pmatrix}
\begin{pmatrix}
k_1 \\
k_2
\end{pmatrix}.
\end{equation}
As we can choose $\psi$ arbitrarily, we can always choose 

\begin{equation}
\tan{2\psi} = -k_2/k_1,
\end{equation}
such that $k_2' = 0$, in which case $\vec{a}'^{R}$ and $\vec{a}'^{I}$ are orthogonal. In addition, we can always choose $k'_1 = k_1/\cos{2\psi}\geq 0 $ such that $|\vec{a}'^{R}| \geq |\vec{a}'^{I}|$. Therefore, via the phase freedom in $\vec{a}$, we can assume, without loss of generality, that $\vec{a}^R\cdot\vec{a}^I = 0$ and that  $|\vec{a}^{R}| \geq |\vec{a}^{I}|$. Taking into account the fact that $|\vec{a}| = 1$, we see that in order to simulate any GKS matrix $\vec{a}\vec{a}^{\dagger}$, it is sufficient to consider only

\begin{align}
\vec{a} &=\vec{a}^{R} + i\vec{a}^{I}\label{realima}\\
 &= \cos{(\theta)}\hat{a}^{R} + i\sin{(\theta)}\hat{a}^{I} \label{trigreal}
\end{align} 
with $|\hat{a}^{R}| = |\hat{a}^{I}| = 1$, $\theta \in [0,\pi/4]$ and $\hat{a}^R\cdot\hat{a}^I = 0$.

Now, as per Theorem \ref{unconj}, we see that conjugation via $U \in \mathrm{SU}(d)$, of the semigroup whose generator is specified by GKS matrix $\vec{a}\vec{a}^{\dagger}$, results in the transformation

\begin{equation}
\vec{a}\vec{a}^{\dagger} \rightarrow G_{(U)}\vec{a}\vec{a}^{\dagger}G_{(U)}^T = (G_{(U)}\vec{a})(G_{(U)}\vec{a})^{\dagger},
\end{equation}
where $G_{(U)} = \hat{\mathrm{Ad}}(U) \in \mathrm{Int}\big(\mathrm{SU}(d)\big) $ is a real matrix. Furthermore, using the natural basis isomorphism $f:\mathrm{su}(d)\rightarrow \mathbb{R}^{d^2-1}$,  we see that

\begin{align}
G_{(U)}\vec{a} &= \cos{(\theta})f\big[\mathrm{Ad}(U)(\hat{A}^R)\big] + i\sin{(\theta)}f\big[\mathrm{Ad}(U)(\hat{A}^I)\big]\nonumber
\\&= \cos{(\theta})f\big[U\hat{A}^RU^{\dagger}\big] + i\sin{(\theta)}f\big[U\hat{A}^IU^{\dagger}\big],
\end{align}
where we have defined $\hat{A}^R \equiv f^{-1}(\hat{a}^{R})$ and $\hat{A}^I\equiv f^{-1}(\hat{a}^{I})$.

At this stage it is useful to define an explicit basis for $\mathrm{su}(d)$. To this end, let $\{|j\rangle\}|_{j = 1}^{d}$ be a basis for $\mathbb{R}^{d}$ and define the Hermitian traceless matrices

\begin{align}
d^{(l)} &= \frac{1}{\sqrt{l(l+1)}}\bigg[\sum_{j = 1}^l|j\rangle\langle j | - l|l+1\rangle\langle l+1|   \bigg], \\
\sigma_x^{(j,k)} &= \frac{1}{\sqrt{2}}\bigg(|j\rangle\langle k| + |k\rangle\langle j|  \bigg), \\
\sigma_y^{(j,k)} &= \frac{1}{\sqrt{2}}\bigg(-i|j\rangle\langle k| + i|k\rangle\langle j|  \bigg), 
\end{align}
such that
\begin{equation}\label{basis}
 \bigg\lbrace \{id^{(l)}\}\big|_{l = 1}^{d-1},  \{i\sigma_x^{(j,k)}, i\sigma_y^{(j,k)}\}\big|_{j = 1}^{d-1}\big|_{j<k\leq d}\bigg\rbrace    
\end{equation}
is a basis for $\mathrm{su}(d)$ and $\{id^{(l)}\}\big|_{l = 1}^{d-1}$ is a basis for the diagonal Cartan subalgebra of $\mathrm{su}(d)$.

As $\hat{A}^{R} \in \mathrm{su}(d)$, we can always find $U_1 \in \mathrm{SU}(d)$ which diagonalises $\hat{A}^{R}$, such that

\begin{equation}\label{first}
U_1\hat{A}^{R}U_1^{\dagger} \equiv \tilde{A}^{R}_d = \sum_{l = 1}^{d-1}d^{R}_{l}(id^{(l)}),
\end{equation}
with real components $\{d^{R}_{l}\}$. Defining $\tilde{A}^{I}  \equiv U_1\hat{A}^{I}U_1^{\dagger}$, we can also write
\begin{align}
\tilde{A}^{I}&\equiv \tilde{A}^{I}_d + \tilde{A}^{I}_{\sigma} \\\label{second}
&=  \sum_{l = 1}^{d-1}d^{I}_{l}(id^{(l)})\nonumber\\ &\quad+ \sum_{j = 1}^{d-1}\sum_{k = j+1}^d\bigg(a^x_{(j,k)}(i\sigma_x^{(j,k)}) + a^y_{(j,k)}(i\sigma_y^{(j,k)})  \bigg),
\end{align}
with real components $\{d^{I}_{l}\}$, $\{a^x_{(j,k)}\}$ and $\{a^y_{(j,k)}\}$. 

Now, let $U_2 = \mathrm{exp}(i\sum_{l = 1}^{d-1}h_ld^{(l)})$ for some $\vec{h} \in \mathbb{R}^{d-1}$ with components $h_l$. One can then see that for any $\vec{h} \in \mathbb{R}^{d-1}$,

\begin{align}
U_2\tilde{A}^{R}_dU_2^{\dagger} &= \tilde{A}^{R}_d, \\
U_2\tilde{A}^{I}_dU_2^{\dagger} &= \tilde{A}^{I}_d,
\end{align}
so that if we define $\tilde{B}^{I}_{\sigma} \equiv U_2\tilde{A}^{I}_\sigma U_2^{\dagger}$ and take $G_{(U_2U_1)} \equiv \hat{\mathrm{Ad}}(U_2U_1)$, then we obtain,

\begin{equation}
G_{(U_2U_1)}\vec{a} = \cos(\theta)f\big(\tilde{A}^{R}_d\big) + i\sin(\theta)f\big( \tilde{A}^{I}_d + \tilde{B}^{I}_{\sigma} \big).
\end{equation} 

In order to obtain an explicit expression for $\tilde{B}^{I}_{\sigma}$ let us  define the matrices $\sigma^{(j,k)} \equiv (1/\sqrt{2})|j\rangle\langle k|$, and rewrite $\tilde{A}^{I}_{\sigma}$ as

\begin{equation}
\tilde{A}^{I}_{\sigma} = \sum_{j = 1}^{d-1}\sum_{k = j+1}^d\bigg(a_{(j,k)}(i\sigma^{(j,k)}) + \overline{a}_{(j,k)}(i\sigma^{(k,j)})  \bigg),
\end{equation}
where 
\begin{equation}
a_{(j,k)} = a^x_{(j,k)} - i a^y_{(j,k)} \equiv m_{(j,k)}e^{i\phi_{(j,k)}},
\end{equation}
and $\overline{a}_{(j,k)}$ denotes the complex conjugate of $a_{(j,k)}$. The matrices $\sigma^{(j,k)}$ are eigenvectors of the map which conjugates by $U_2$, so that some algebra yields,

\begin{equation}
\tilde{B}^{I}_{\sigma} = \sum_{j = 1}^{d-1}\sum_{k = j+1}^d i\bigg[m_{(j,k)}e^{if_{(j,k)}}(\sigma^{(j,k)}) + \mathrm{H.C} \bigg], 
\end{equation}
where $\mathrm{H.C}$ denotes the Hermitian conjugate, and
\begin{align}
f_{(j,k)} &= \phi_{(j,k)} - (j-1)\varphi(j-1)h_{j-1} \nonumber\\ &\qquad + \sum_{l = j}^{k-2}\big(\varphi(l)h_{l}\big) + k\varphi(k-1)h_{k-1} ,
\end{align}
with $\varphi(j) \equiv 1/(\sqrt{j(j+1)})$ and $h_0 \equiv 0$. If we choose

\begin{equation}
h_1 = -\frac{1}{2\varphi(1)}\phi_{(1,2)},
\end{equation}
and then inductively set

\begin{equation}
h_l = -\frac{1}{(l+1)\varphi(l)}\big[\phi_{1,l+1} + \sum_{x = 1}^{l-1}\varphi(x)h_x   \big],
\end{equation}
we see that $f_{(1,k)} = 0$ for all $k \in [2,d]$. As a result, we obtain that

\begin{align}
\tilde{B}^{I}_{\sigma} &= \sum_{k = 2}^d m_{(1,k)}i(\sigma^{(1,k)} + \sigma^{(k,1)})\nonumber\\&\qquad + \sum_{j = 2}^{d-1}\sum_{k = j+1}^d i\bigg[m_{(j,k)}e^{if_{(j,k)}}(\sigma^{(j,k)}) + \mathrm{H.C}\bigg] \\
&= \sum_{k = 2}^d m_{(1,k)}(i\sigma^{(1,k)}_x) \nonumber\\
&\qquad + \sum_{j = 2}^{d-1}\sum_{k = j+1}^d\bigg(b^x_{(j,k)}(i\sigma_x^{(j,k)}) + b^y_{(j,k)}(i\sigma_y^{(j,k)})  \bigg).\label{third}
\end{align}
If we now define $\tilde{a}^R = f(\tilde{A}^R_d)$ and $\tilde{a}^I = f(\tilde{A}^I_d + \tilde{B}^{I}_{\sigma})$, then by fixing an appropriate order for the basis vectors in \eqref{basis}, and relabelling the components in \eqref{first}, \eqref{second}, \eqref{third}, we can write

\begin{equation}\label{form}
\tilde{a}^R = \begin{pmatrix}
a^R_1 \\
\vdots \\
a^R_{d-1}\\
0\\
\vdots\\
\vdots\\
0
\end{pmatrix} \qquad \tilde{a}^I = \begin{pmatrix}
a^I_1 \\
\vdots \\
\vdots\\
a^R_{d^2-d}\\
0\\
\vdots\\
0
\end{pmatrix}.
\end{equation}

Furthermore, via complete antisymmetry of the structure constants of $\mathrm{su}(d)$ one can prove that $\mathrm{Int}(\mathrm{SU}(d)) \subseteq \mathrm{SO}(d^2-1)$, and therefore that the adjoint action preserves orthogonality and normalisation. As a result, we have now successfully shown that for any GKS matrix $\vec{a}\vec{a}^{\dagger}$, with $\vec{a} \in \mathbb{C}^{d^2-1}$ and $|\vec{a}| = 1$, there always exists $U = U_2U_1 \in \mathrm{SU}(d)$ such that 

\begin{equation}
G_{(U)}\vec{a} = \cos(\theta)\tilde{a}^R + i\sin(\theta)\tilde{a}^I,
\end{equation} 
where $G_{(U)} = \hat{\mathrm{Ad}}(U)$ and $\tilde{a}^R, \tilde{a}^I \in \mathbb{R}^{d^2-1}$ are given by \eqref{form}, with $|\tilde{a}^R| = |\tilde{a}^I| = 1$ and $\tilde{a}^R.\tilde{a}^I = 0$. Exploiting orthogonality and normalisation we can always find angles $\{\alpha^R_j\}|_{j = 1}^{d-2}$ and $\{\alpha^I_k\}|_{k = 1}^{d^2 - (d+1)}$ such that the parametrisation given in the statement of the theorem exists. Finally, using the definition of  $G_{(U)}$, along with complete antisymmetry of the structure constants, one can show that $G_{(U)}^T = G_{(U^{\dagger})} = \hat{\mathrm{Ad}}(U^\dagger) $, and therefore as $G_{(U)} \in \mathrm{SO}(d^2-1)$ we have that

\begin{equation}
\vec{a}\vec{a}^{\dagger} = G_{(U^{\dagger})}\big[G_{(U)} \vec{a}\vec{a}^{\dagger}  G_{(U)}^T   \big]G_{(U)^\dagger}^T,
\end{equation}
and as a result the semigroup generated by $\vec{a}\vec{a}^{\dagger}$ can be simulated through the semigroup generated by $G_{(U)} \vec{a}\vec{a}^{\dagger}  G_{(U)}^T $, a member of the universal set, using unitary conjugation via $U^{\dagger}$. \\

\subsection{Proof of necessity} We show here that using linear combination and unitary conjugation it is not possible to simulate the Markovian semigroup specified by some GKS matrix $A(\theta,\vec{\alpha}^R,\vec{\alpha}^I)$, satisfying the restrictions of the theorem statement, through simulation of some other combination/transformation of Markovian semigroups specified by GKS matrices satisfying the same conditions for some different set of parameters. 

Firstly, all $A(\theta,\vec{\alpha}^R,\vec{\alpha}^I)$, as projections onto the eigenspace of a single eigenvector of $A$, a basis vector of $\mathbb{C}^{d^2-1}$, are rank one matrices. As rank one matrices are extreme in the convex cone of positive matrices, no such $A(\theta,\vec{\alpha}^R,\vec{\alpha}^I)$ can be simulated through the linear combination of Markovian semigroups specified by other such GKS matrices. 
Note also that a phase transformation of $\vec{a}(\theta,\vec{\alpha}^R,\vec{\alpha}^{I})$ commutes with a rotation via $G \in \mathrm{Int}(\mathrm{SU}(d))$, and as such we only need to prove that if $\vec{a}(\theta,\vec{\alpha}^R,\vec{\alpha}^{I})$ and $\vec{a}(\theta',\vec{\alpha}'^R,\vec{\alpha}'^{I})$ satisfy the restrictions \eqref{firstone}-\eqref{lastone}, but for different sets of parameters, and

\begin{equation}
e^{i\psi}G \big[\vec{a}(\theta,\vec{\alpha}^R,\vec{\alpha}^{I})\big] = \vec{a}(\theta',\vec{\alpha}'^R,\vec{\alpha}'^{I}), \label{condition}\\
\end{equation}
for some $\psi \in [0,2\pi]$ and some $G \in \mathrm{Int}(\mathrm{SU}(d))$, then $(\theta,\vec{\alpha}^R,\vec{\alpha}^{I}) = (\theta',\vec{\alpha}'^R,\vec{\alpha}'^{I})$.  In order to simplify the presentation of the proof, in what follows we drop from our notation the explicit dependency of vectors on their parameters by defining

\begin{align}
 \vec{a}(\theta,\vec{\alpha}^R,\vec{\alpha}^{I}) &=  \cos(\theta)\tilde{a}^R(\vec{\alpha}^R) + i\sin(\theta)\tilde{a}^I(\vec{\alpha}^{I})\\
 &\equiv \cos(\theta)\tilde{a}^R + i\sin(\theta)\tilde{a}^I\\
 &\equiv \vec{a}^R + \vec{a}^I\\
 &\equiv \vec{a},
\end{align}
and
\begin{align}
 \vec{a}(\theta',\vec{\alpha}'^R,\vec{\alpha}'^{I}) &=  \cos(\theta')\tilde{a}^R(\vec{\alpha}'^R) + i\sin(\theta')\tilde{a}'^I(\vec{\alpha}'^{I})\\
 &\equiv \cos(\theta')\tilde{a}'^R + i\sin(\theta')\tilde{a}'^I\\
 &\equiv \vec{a}'^R + \vec{a}'^I\\
 &\equiv \vec{a}',
\end{align}
with the goal of proving that if $e^{i\psi}G\vec{a} = \vec{a}'$ then $\vec{a} = \vec{a}'$. In this simplified notation we can write,

\begin{equation}
e^{i\psi}G \big[\vec{a}\big]
= e^{i\psi} \big[\cos(\theta)\big(G\tilde{a}^R\big) + i\sin(\theta)\big(G\tilde{a}^I\big) \big],
\end{equation}
where, as $G \in \mathrm{Int}(\mathrm{SU}(d)) \subseteq \mathrm{SO}(d^2-1)$, we see that rotation of $\vec{a}(\theta,\vec{\alpha}^R,\vec{\alpha}^{I})$ via $G$ leaves $\theta$ unchanged. Furthermore, if we define

\begin{align}
\tilde{k}_1 &\equiv |\cos(\theta)\big(G\hat{a}^R\big) |^2 - |\sin(\theta)\big(G\hat{a}^I\big)|^2 \\
\tilde{k}_2 &\equiv 2 \big[\cos(\theta)\big(G\hat{a}^R\big)]\cdot\big[\sin(\theta)\big(G\hat{a}^I\big)\big]  
\end{align}
then via the fact that $G \in \mathrm{SO}(d^2-1)$ we obtain that $\tilde{k}_1 = k_1 \geq 0 $ and $\tilde{k}_2 = k_2 = 0 $, where $k_1$ and $k_2$ are defined in \eqref{k1} and \eqref{k2}. Let us now define

\begin{align}
k'_1 &\equiv |\vec{a}'^R|^2 - |\vec{a}'^I|^2 \\
k'_2 &\equiv 2 \vec{a}'^R\cdot\vec{a}'^I. 
\end{align}
$(k'_1,k'_2)$ is related to $(\tilde{k}_1,\tilde{k}_2)$ via an expression such as \eqref{transformation}, but as $k'_1 \geq 0$ and $k'_2 = 0$ by assumption, we see that we must have $\psi = 0$, i.e. the phase transformation must be trivial. As neither the phase transformation nor rotation via $G$ effects $\theta$, we have that $\theta = \theta'$ and we can then write

\begin{align}
\vec{a}' 
&= \cos(\theta)\big(G\tilde{a}^R\big) + i\sin(\theta)\big(G\tilde{a}^I)\\
&= \cos(\theta)\tilde{a}'^R + i\sin{\theta}\tilde{a}'^I.
\end{align}
Furthermore, again because $G \in \mathrm{Int}(\mathrm{SU}(d)) \subseteq \mathrm{SO}(d^2-1)$ is real, $e^{i\psi}G\vec{a} = \vec{a}'$ implies that $G\tilde{a}^R  = \tilde{a}'^R$ and $G\tilde{a}^I = \tilde{a}'^I$, and therefore all that remains is to prove that $G\tilde{a}^R = \tilde{a}^R$ and $G\tilde{a}^I = \tilde{a}^I$.

To this end, let us define

\begin{align}
\tilde{A}^R &= f^{-1}(\tilde{a}^R), \\
\tilde{A}'^R &= f^{-1}(\tilde{a}'^R). 
\end{align}
If $G = \hat{\mathrm{Ad}}(U)$, for some $U \in \mathrm{SU}(d)$, then from $G\tilde{a}^R  = \tilde{a}'^R$ we have that 

\begin{equation}
\tilde{A}'^R = U\tilde{A}^RU^{\dagger}.
\end{equation}
However, also by assumption, both $\tilde{A}'^R$ and $\tilde{A}^R$ are diagonal, and therefore $U$ must also be diagonal, i.e. we must have that $U = \mathrm{exp}(i\sum_{l = 1}^{d-1}p_ld^{(l)})$, for some $\vec{p} \in \mathbb{R}^{d-1}$ with components $p_l$. However, in this case one can show that 

\begin{equation}
U\tilde{A}^RU^{\dagger} = \tilde{A}^R,
\end{equation}
and therefore $G\tilde{a}^R = \tilde{a}^R$. To prove that $G\tilde{a}^I = \tilde{a}^I$ we define

\begin{equation}
f^{-1}(\tilde{a}^I) \equiv \tilde{A}^I_d + \tilde{B}^{I}_{\sigma} ,
\end{equation}
where

\begin{equation}
\tilde{A}^I_d = \sum_{l = 1}^{d-1}d^{I}_{l}(id^{(l)}),
\end{equation}
is diagonal and 

\begin{align}
\tilde{B}^{I}_{\sigma} &= \sum_{k = 2}^d m_{(1,k)}i(\sigma^{(1,k)} + \sigma^{(k,1)})\nonumber\\&\qquad + \sum_{j = 2}^{d-1}\sum_{k = j+1}^d i\bigg[m_{(j,k)}e^{if_{(j,k)}}(\sigma^{(j,k)}) + \mathrm{H.C}\bigg]. \label{zeroexp}
\end{align}
We then have that

\begin{equation}\label{happening}
G\tilde{a}^I = f(U\tilde{A}^I_dU^{\dagger} + U\tilde{B}^I_{\sigma}U^\dagger),
\end{equation}
but given diagonal $U$ we again see that

\begin{equation}\label{diaghand}
U\tilde{A}^I_dU^{\dagger} = \tilde{A}^I_d,
\end{equation}
and as such all that remains is to prove that $U\tilde{B}^I_{\sigma}U^{\dagger} = \tilde{B}^I_{\sigma}$. To show this, note that via our assumptions we can write

\begin{equation}\label{tildeform}
f^{-1}(\tilde{a}'^I) \equiv \tilde{A}'^I_d + \tilde{B}'^{I}_{\sigma},
\end{equation}
with $\tilde{A}'^I_d$ diagonal and 

\begin{align}
\tilde{B}'^{I}_{\sigma} &= \sum_{k = 2}^d m'_{(1,k)}i(\sigma^{(1,k)} + \sigma^{(k,1)})\nonumber\\&\qquad + \sum_{j = 2}^{d-1}\sum_{k = j+1}^d i\bigg[m'_{(j,k)}e^{if'_{(j,k)}}(\sigma^{(j,k)}) + \mathrm{H.C}\bigg].\label{correctform}
\end{align}
However, from \eqref{happening}-\eqref{tildeform} and the fact that $G\tilde{a}^I  = \tilde{a}'^I$, we can also see that 

\begin{equation}
\tilde{B}'^{I}_{\sigma} = U\tilde{B}^{I}_{\sigma}U^{\dagger},
\end{equation}
and therefore that

\begin{equation}\label{comparison}
\tilde{B}'^{I}_{\sigma} = \sum_{j = 1}^{d-1}\sum_{k = j+1}^d i\bigg[m_{(j,k)}e^{i\gamma_{(j,k)}}(\sigma^{(j,k)}) + \mathrm{H.C} \bigg], 
\end{equation}
where

\begin{align}\label{equations}
\gamma_{(j,k)} &= f_{(j,k)} - (j-1)\varphi(j-1)p_{j-1} \nonumber\\ &\qquad + \sum_{l = j}^{k-2}\big(\varphi(l)p_{l}\big) + k\varphi(k-1)p_{k-1},
\end{align}
and from \eqref{zeroexp} we have that $f_{(1,k)} = 0$ for $k \in [2,d]$. By comparison of \eqref{correctform} and \eqref{comparison} we see that we must have $\gamma_{(1,k)}  = 0$ for $k \in [2,d]$, and therefore from \eqref{equations} and \eqref{zeroexp} we can show that we must have $p_l = 0$ for $l \in [1,d-1]$. This implies that $U = \mathds{1}$, and therefore $G = \mathds{1}$ and $G\vec{a} = \vec{a}$. $\blacksquare$

\section{Simulation Algorithm}\label{simsec}

If we assume the ability to implement the necessary and sufficient set of resources implied by Theorem \ref{main}, or in other words, if we consider arbitrary unitary operations on our system along with quantum channels from the semigroups generated by elements of the universal set as ``standard resources", then the construction of Theorem  \ref{main}, along with previous work on simulation of linear combinations \cite{Sweke2014} (described in Appendix A), implies a natural algorithm for the simulation of arbitrary Markovian open quantum systems. This algorithm is not necessarily efficient for an arbitrary system, however after presentation of the algorithm we discuss the conditions under which a Markovian open quantum system can be efficiently simulated, with respect to the constructed universal set, using this algorithm. This discussion of efficiency relies on the details concerning simulation of linear combinations \cite{Sweke2014}, as presented in detail in Appendix \ref{lincomb}.

The algorithm is as follows:

\begin{enumerate}
\item Given $\mathcal{H}_S \cong \mathbb{C}^d$ and $\mathcal{L} = \mathcal{L}_H + \mathcal{L}_A \in \mathcal{B}(\mathcal{B}(\mathcal{H}_S))$, the generator of a Markovian semigroup, obtain the spectral decomposition of $A$ such that

\begin{equation}\label{lincombequation}
\mathcal{L} = \mathcal{L}_H + \sum_{k=1}^{m}\lambda_k\mathcal{L}_{\vec{a}_{k}\vec{a}^{\dagger}_{k}} \equiv \sum_{k = 0}^{m}\lambda_k\mathcal{L}_k,
\end{equation} 
where $\lambda_0 = 1$ and $\mathcal{L}_0 \equiv \mathcal{L}_H$. 
\item For each $k \in [1,m]$ use phase freedom to find $\theta_k$, and construct $U^{(k)}_1$ and $U^{(k)}_2$ as per the proof of Theorem \ref{main}, such that by defining $U^{(k)} = U^{(k)\dagger}_1U^{(k)\dagger}_2 $,

\begin{equation}
\vec{a}_{k}\vec{a}^{\dagger}_k = G_{(U^{(k)})}\big[A^{(k)}(\theta_k,\vec{\alpha}_k^R,\vec{\alpha}_k^{I})\big]G_{(U^{(k)})}^T,
\end{equation}
where $A^{(k)}(\theta_k,\vec{\alpha}_k^R,\vec{\alpha}_k^{I})$ is an element of the universal set of semigroup generators.

\item Given $\epsilon > 0$ and $t>0$, construct, as described in Appendix \ref{lincomb}, the Suzuki first-order integrator $S_{2}(\hat{\mathcal{L}}_1,\ldots,\hat{\mathcal{L}}_m,t/r)$ \cite{Sweke2014, Papageorgiou2012}, with

\begin{equation}
r =  \frac{\sqrt{2L_2}(mt)^{3/2}}{\epsilon^{1/2}},
\end{equation}
where $L_2 := ||\mathcal{L}_2||_{1\rightarrow 1}$.

\item Given $\rho(0)$, implement $S_{2}(\hat{\mathcal{L}}_1,\ldots,\hat{\mathcal{L}}_m,t/r)$ consecutively $rL_1$ times, in order to recombine the linear combination \eqref{lincombequation} through sequential implementations of $T_k(\tilde{t}) = e^{\tilde{t}\mathcal{L}_k}$. Each implementation of $T_k(\tilde{t})$ is achieved via 

\begin{equation}\label{threeresources}
T_k(\tilde{t}) = \mathcal{U}^{\dagger}_{k}\big(T_{A^{(k)}}(\tilde{t})\big)\mathcal{U}_k, 
\end{equation}
where $\mathcal{U}_k(\rho) = U^{(k)}(\rho)U^{(k)\dagger}$ and $T_{A^{(k)}}(t) = \mathrm{exp}(t\mathcal{L}_{A^{(k)}})$.

\end{enumerate}

As shown in \cite{Sweke2014}, and presented in Appendix \ref{lincomb},  as a result of the Suzuki-Lie-Trotter procedure used for the recombination of linear combinations, the above algorithm simulates the Markovian semigroup generated by \eqref{lincombequation}, within precision $\epsilon$, using $\mathrm{poly}\big(||\mathcal{L}||_{(1\rightarrow 1)},t,1/\epsilon, m \big)$ applications of ``standard resources", i.e. implementations of quantum channels from the semigroups generated by elements of the universal set and unitary operations on the system. More precisely, the algorithm requires at most 

\begin{equation}\label{res}
\mathrm{N} \leq (2m-1)  \frac{\sqrt{2L_2}L_1(mt)^{3/2}}{\epsilon^{1/2}}.
\end{equation}
implementations of channels $T_k(\tilde{t})$, each of which, as per Eq. \eqref{threeresources}, requires $3$ ``standard resources", namely two unitary operations and one quantum channel from a semigroup generated by an element of the universal set.

 By comparison with our definition of \textit{efficient} simulation in Section \ref{setting}, we therefore see that this algorithm will be efficient, with respect to this universal set, for any class of Markovian semigroups for which $m$, the number of non-zero eigenvalues of the GKS matrix $A$, is proportional to $\mathrm{ln}(d)$, or alternatively, if we are within a many-body context, to the number of subsystems. As $A \in \mathcal{M}_{d^2-1}(\mathbb{C}) $, we see that in the general case $m = d^2-1$ and the algorithm will not be efficient - however by comparing the GKSL form of Eq. \eqref{GKS} with the diagonal form of Eq. \eqref{diagonalform} we see that the algorithm will be efficient, with respect to this universal set, for any system for which the number of distinct physical dissipation processes with non-zero rates (the number of distinct Lindblad operators) scales polynomially with the number of subsystems. 
 
 \begin{figure} 
 \includegraphics[scale = 0.4]{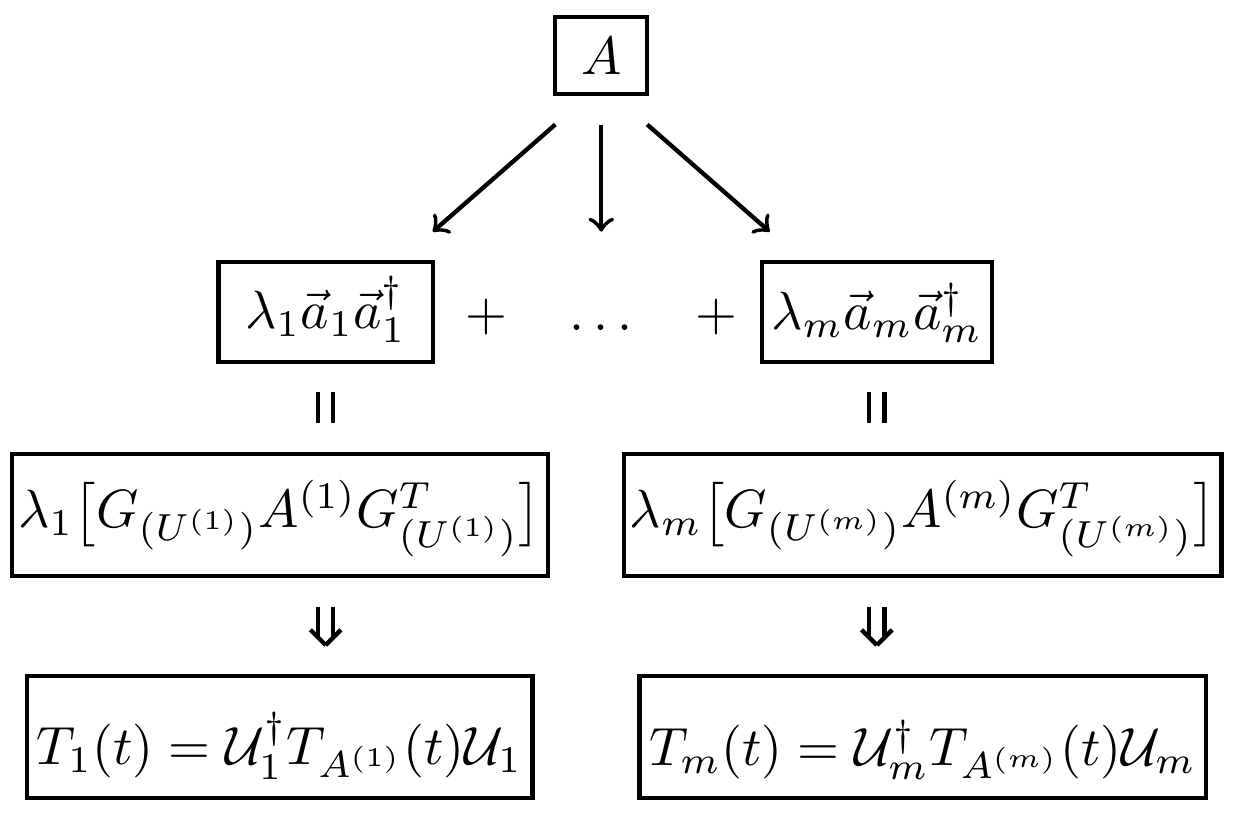} 
 \caption{Any GKS matrix $A \geq 0$ can be decomposed into the linear combination of rank $1$ GKS matrices $\vec{a}_i\vec{a}_i^{\dagger}$. The semigroups whose generator is specified by these matrices can be further decomposed into the unitary conjugation of semigroups whose generator is specified by an element of the universal set $A^{(i)}(\theta_i,\vec{\alpha}_i^R,\vec{\alpha}_i^{I})$. As a result any quantum channel from the original semigroup can be implemented through the linear combination and unitary conjugation of channels from the semigroups whose generators belong to the universal set.}  \label{operatoraction}
\end{figure}

\section{Worked Example}

\begin{figure} 
 \includegraphics[scale = 1]{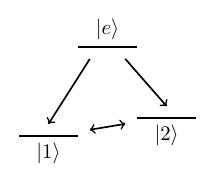} 
 \caption{Illustration of three level $\Lambda$ atom experiencing effective collective spontaneous emission and external incoherent driving.} \label{threelevel}
\end{figure}

As an illustration of the above algorithm we consider as an example a three level atom in the $\Lambda$ configuration (see Fig. \ref{threelevel}), experiencing effective dissipation described by the Lindblad master equation 

\begin{equation}
\mathcal{L}(\rho) = \sum_{i = 1}^2\gamma_i\bigg(L_i\rho L_i^{\dagger} - \frac{1}{2}\lbrace L_i^{\dagger}L_i,\rho   \rbrace_+   \bigg),
\end{equation}
where,

\begin{align}
L_1 &= \cos\phi |1\rangle\langle e| + e^{i\eta}\sin\phi|2\rangle\langle e|,\\
L_2 &=  \cos\alpha |1\rangle\langle 2| + \sin\alpha|2\rangle\langle 1|.
\end{align}
We begin by transforming into the GKS form,

\begin{equation}
\mathcal{L}_A(\rho) = \sum_{l,k = 1}^{8}A_{l,k}\bigg(F_l\rho F_k^{\dagger} - \frac{1}{2}\lbrace F_k^{\dagger}F_l,\rho   \rbrace_+   \bigg),
\end{equation}
where $\{F_i\}|_{i = 1}^8$ is a Hermitian basis for the traceless matrices in $\mathcal{M}_3(\mathbb{C})$, defined via

\begin{align}
\{F_i\}|_{i = 1}^2 &\equiv \{d^{(l)}\}|_{l = 1}^{2} \\ 
\{F_i\}|_{i = 3}^5 &\equiv \{\sigma_x^{(j,k)}\}_{j = 1}^{2}|_{j<k\leq 3} \\ 
\{F_i\}|_{i = 6}^8 &\equiv \{\sigma_y^{(j,k)}\}_{j = 1}^{2}|_{j<k\leq 3}. 
\end{align} 
Setting $\phi = \eta = \alpha = \pi/3$, we find that with respect to this basis 

\begin{equation}
A = \begin{pmatrix}
0 &0&0 &0&0 &0&0 &0\\
0 &0&0 &0&0 &0&0 &0\\
0 &0&a_{3,3} &a_{3,4}&0 &ia_{3,3}&a_{3,7} &0\\
0 &0&\overline{a_{3,4}} &3a_{3,3}&0 &a_{4,6}&3ia_{3,3} &0\\
0 &0&0 &0&a_{5,5} &0&0 &a_{5,8}\\
0 &0&-ia_{3,3} &\overline{a_{4,6}}&0 &a_{3,3}&a_{3,4} &0\\
0 &0&\overline{a_{3,7}} &-3ia_{3,3}&0&\overline{a_{3,4}} &3a_{3,3} &0\\
0 &0&0 &0&\overline{a_{5,8}} &0&0 &a_{8,8}
\end{pmatrix},
\end{equation}
where the overbar is used to denote the complex conjugate, and

\begin{align}
a_{3,3} &= \frac{\gamma_1}{8} \\
a_{3,4} &= \frac{\sqrt{3} - 3i}{16}\gamma_1\\
a_{3,7} &= \frac{3+i\sqrt{3}}{16}\gamma_1\\
a_{4,6} &= \frac{-3+i\sqrt{3}}{16}\gamma_1\\ 
a_{5,5} &=  \frac{2+\sqrt{3}}{4}\gamma_2\\
a_{5,8} &= \frac{i\gamma_2}{4}\\
a_{8,8} &= \frac{2-\sqrt{3}}{4}\gamma_2.
\end{align}
The next step is to decompose $A$ into the linear combination of rank $1$ generators through the spectral decomposition. Constructing this decomposition we obtain

\begin{equation}
A = \sum_{k = 1}^2\lambda_k\vec{a}_k\vec{a}_k^{\dagger},
\end{equation}
where $\lambda_i = \gamma_i$, and

\begin{equation}\label{a1}
\vec{a}_1 = \frac{\sqrt{3}}{2\sqrt{2}}\begin{pmatrix}
1\\
0 \\
\frac{1}{2}+\frac{i}{2 \sqrt{3}}\\
i\\
0\\
-\frac{1}{2} i+\frac{1}{2 \sqrt{3}}\\
1\\
0
\end{pmatrix}, \end{equation}\begin{equation}\label{secondvec}
 \vec{a}_2 = \frac{1}{\sqrt{1+\left(2+\sqrt{3}\right)^2}}\begin{pmatrix}
0\\
0 \\
0\\
0\\
(2+\sqrt{3})i\\
0\\
0\\
1
\end{pmatrix}.
\end{equation}

At this stage each constituent generator $\vec{a}_i\vec{a}_i^{\dagger}$ of the linear combination needs to be decomposed into the unitary conjugation of a semigroup from the universal set. We focus first on decomposing the semigroup generated by $\vec{a}_1\vec{a}_1^{\dagger}$. The first step in this regard is to identify the phase $\psi_1$ such that 

\begin{equation}
e^{i\psi_1}\vec{a}_1 = \cos(\theta_1)\hat{a}_1^R + i\sin(\theta_1)\hat{a}_1^I,
\end{equation} 
for some $\hat{a}_1^R$ and $\hat{a}_1^I$ such that $\hat{a}_1^R\cdot\hat{a}_1^I = 0$, $|\hat{a}_1^R| = |\hat{a}_1^I| = 1$ and $\theta_1 \in [0,\pi/4]$. For $\vec{a}_1$ as per \eqref{a1} no such phase transformation is necessary, (i.e. we use $\psi_1 = 0$) and we see that

\begin{align}
\vec{a}_1 & = \frac{1}{\sqrt{2}}\begin{pmatrix}
0\\
0 \\
\frac{\sqrt{3}}{4}\\
0\\
0\\
\frac{1}{4}\\
\frac{\sqrt{3}}{2}\\
0
\end{pmatrix} +\frac{1}{\sqrt{2}}i\begin{pmatrix}
0\\
0 \\
\frac{1}{4}\\
\frac{\sqrt{3}}{2}\\
0\\
-\frac{\sqrt{3}}{4}\\
0\\
0
\end{pmatrix}\label{ahat1} \\
& = \cos(\theta_1)\hat{a}_1^R + \sin(\theta_1)\hat{a}_1^I,
\end{align}
with $\theta_1 = \pi/4$. The next step is to identify $U^{(1)}_1$ and  $U^{(1)}_2$ such that

\begin{align}
\tilde{a}_1^R &\equiv f\big(U^{(1)}_2U^{(1)}_1f^{-1}[\hat{a}^{R}_1]U^{(1)\dagger}_1U^{(1)\dagger}_2   \big), \\
\tilde{a}_1^I &\equiv f\big(U^{(1)}_2U^{(1)}_1f^{-1}[\hat{a}^{I}_1]U^{(1)\dagger}_1U^{(1)\dagger}_2  \big), 
\end{align}
have the form given in \eqref{firstform}, where $f:\mathrm{su}(d) \rightarrow \mathbb{R}^{8}$ is the natural isomorphism defined via $f(iF_j) = |j\rangle$, where $\{iF_j\}|_{j = 1}^8$ and $\{|j\rangle\}|_{j = 1}^{8}$ are the standard bases for $\mathrm{su}(d)$ and $\mathbb{R}^{8}$ respectively. As per the proof of Theorem \ref{main}, $U^{(1)}_1$ is the matrix which diagonalises $\hat{A}^R_1 \equiv f^{-1}(\hat{a}^R_1)$. For $\hat{a}^{R}_1$ as per \eqref{ahat1} we find that

\begin{equation}
\hat{A}^R_1 = \frac{\sqrt{3}}{2\sqrt{2}}\begin{pmatrix}
 0 & \frac{1}{6} \left(3 i+\sqrt{3}\right) & 1 \\
 \frac{1}{6} \left(3 i-\sqrt{3}\right) & 0 & 0 \\
 -1 & 0 & 0
\end{pmatrix},
\end{equation}
and

\begin{equation}
U^{(1)}_1 = \frac{\sqrt{3}}{2\sqrt{2}}\begin{pmatrix}
 \frac{2 i}{\sqrt{3}} & \frac{1}{6} \left(3 i+\sqrt{3}\right) & 1 \\
 -\frac{2 i}{\sqrt{3}} & \frac{1}{6} \left(3 i+\sqrt{3}\right) & 1 \\
 0 & \sqrt{2} \left(-\frac{1}{2}-\frac{i \sqrt{3}}{2}\right) & \sqrt{\frac{2}{3}}
\end{pmatrix}
\end{equation}
such that

\begin{equation}
\tilde{A}^{R}_{d,1} \equiv  U^{(1)}_1\hat{A}^{R}_1U^{(1)\dagger}_1 
= \begin{pmatrix}
 \frac{i}{\sqrt{2}} & 0 & 0 \\
 0 & -\frac{i}{\sqrt{2}} & 0 \\
 0 & 0 & 0
\end{pmatrix}
\end{equation}
and
\begin{equation}
{A}^{I}_{1} \equiv  U^{(1)}_1\hat{A}^{I}_1U^{(1)\dagger}_1 
= \begin{pmatrix}
 0 & -\frac{1}{\sqrt{2}} & 0 \\
 \frac{1}{\sqrt{2}} & 0 & 0 \\
 0 & 0 & 0
\end{pmatrix}.
\end{equation}

At this stage one would typically construct diagonal $U^{(1)}_2$ to eliminate $2$ (i.e. $d-1$ with $d = 3$) components of $f(\tilde{A}^{I}_1)$ while leaving $f(\tilde{A}^{R}_{d,1})$ unchanged. However, in this case we see that 

\begin{equation}
f(\tilde{A}^{R}_{d,1}) = \begin{pmatrix}
 1 \\
 0 \\
 0 \\
 0 \\
 0 \\
 0 \\
 0 \\
 0
\end{pmatrix} \quad f(\tilde{A}^{I}_1) = \begin{pmatrix}
 0 \\
 0 \\
 0 \\
 0 \\
 0 \\
 -1 \\
 0 \\
 0
\end{pmatrix},
\end{equation}
so that if we define $\tilde{a}^{R}_1 \equiv f(\tilde{A}^{R}_{d,1})$ and $\tilde{a}^{I}_1 \equiv f(\tilde{A}^{I}_{1})$ then a second unitary transformation is not necessary, as $\tilde{a}^{R}_1$ and $\tilde{a}^{I}_1$ already have the desired form. So, following the proof by defining $U^{(1)} = U^{(1)\dagger}_1$ and $G_{(U^{(1)})} = \hat{\mathrm{Ad}}(U^{(1)})$,  we now have that 

\begin{equation}\label{conjugationone}
\vec{a}_{1}\vec{a}^{\dagger}_1 = G_{(U^{(1)})}\big[A^{(1)}(\theta_1,\vec{\alpha}_1^R,\vec{\alpha}_1^{I})\big]G_{(U^{(1)})}^T,
\end{equation}
where $A^{(1)}(\theta_1,\vec{\alpha}_1^R,\vec{\alpha}_1^{I}$ is an element of the universal set of semigroup generators, with $\theta_1 = \pi/4$, $\vec{\alpha}_1^R = 0$ and 

\begin{equation}
 \vec {\alpha}_1^I = \frac{\pi}{2}\begin{pmatrix}
 1 \\
 1 \\
 1 \\
 1 \\
 3 \\

\end{pmatrix}.
\end{equation}
Furthermore, from Theorem \ref{unconj} and \eqref{conjugationone}, one has that for any channel $T_1(t) = \mathrm{exp}(t\mathcal{L}_{\vec{a}_1\vec{a}_1^{\dagger}})$ from the semigroup generated by $\vec{a}_1\vec{a}_1^{\dagger}$,

\begin{equation}\label{chan1}
T_1(t)(\rho) =    U^{(1)\dagger}\bigg(T_{A^{(1)}}(t)\big[U^{(1)}\rho U^{(1)\dagger}\big]\bigg)U^{(1)},
\end{equation}
where $T_{A^{(k)}}(t) = \mathrm{exp}(t\mathcal{L}_{A^{(k)}})$.

We can now proceed to decompose the semigroup generated by $\vec{a}_2\vec{a}_2^{\dagger}$, the second component of the linear decomposition. We follow the same procedure, however in this case if we simply rewrite \eqref{secondvec} as

\begin{equation}
\vec{a}_2 = \vec{a}^{R}_2 + i\vec{a}^{I}_2 
\end{equation}
then we see that although $\vec{a}^{R}_2\cdot\vec{a}^{I}_2 = 0$, we have $|\vec{a}^{I}_2|> |\vec{a}^{R}_2|$, and as such a non-trivial phase transformation is necessary in order to be able to write

\begin{equation}\label{desireform}
e^{i\psi_2}\vec{a}_2 = \cos(\theta_2)\hat{a}_2^R + i\sin(\theta_2)\hat{a}_2^I,
\end{equation} 
for some $\hat{a}_2^R$ and $\hat{a}_2^I$ such that $\hat{a}_2^R\cdot\hat{a}_2^I = 0$, $|\hat{a}_2^R| = |\hat{a}_2^I| = 1$ and $\theta_2 \in [0,\pi/4]$. As $\vec{a}^{R}_2\cdot\vec{a}^{I}_2 = 0$ we see that a phase transformation via $\psi_2 = \pi/2$ is sufficient, and after such a transformation we obtain an expression in the form \eqref{desireform} with

\begin{equation}
\theta_2 = \arccos\Big(\frac{2+\sqrt{3}}{\sqrt{1+\left(2+\sqrt{3}\right)^2}}\Big),
\end{equation}

and 

\begin{equation}
\hat{a}_2^R = \begin{pmatrix}
 0 \\
 0 \\
 0 \\
 0 \\
 -1 \\
 0 \\
 0 \\
 0
\end{pmatrix} \quad \hat{a}_2^I = \begin{pmatrix}
 0 \\
 0 \\
 0 \\
 0 \\
 0 \\
 0 \\
 0 \\
 1
\end{pmatrix}
\end{equation}
Once again, the next step is to find the unitary matrix $U^{(2)}_1$ which diagonalises $\hat{A}^{R}_2 \equiv f^{-1}(\hat{a}^{R}_2)$. In this case we find

\begin{equation}
\hat{A}^{R}_2 = \begin{pmatrix}
 0 & 0 & 0 \\
 0 & 0 & -\frac{i}{\sqrt{2}} \\
 0 & -\frac{i}{\sqrt{2}} & 0
\end{pmatrix},
\end{equation}
and

\begin{equation}
U^{(2)}_1 = \begin{pmatrix}
 0 & -\frac{1}{\sqrt{2}} & \frac{1}{\sqrt{2}} \\
 0 & \frac{1}{\sqrt{2}} & \frac{1}{\sqrt{2}} \\
 1 & 0 & 0
\end{pmatrix},
\end{equation}
such that
\begin{equation}\label{A2R}
\tilde{A}^{R}_{d,2} \equiv  U^{(2)}_1\hat{A}^{R}_2U^{(2)\dagger}_1 
= \begin{pmatrix}
 \frac{i}{\sqrt{2}} & 0 & 0 \\
 0 & -\frac{i}{\sqrt{2}} & 0 \\
 0 & 0 & 0
\end{pmatrix},
\end{equation}
and

\begin{equation}\label{A2I}
{A}^{I}_{2} \equiv  U^{(2)}_1\hat{A}^{I}_2U^{(2)\dagger}_1 
= \begin{pmatrix}
 0 & -\frac{1}{\sqrt{2}} & 0 \\
 \frac{1}{\sqrt{2}} & 0 & 0 \\
 0 & 0 & 0
\end{pmatrix}.
\end{equation}
From \eqref{A2R} and \eqref{A2I} we see that $\tilde{A}^{R}_{d,2} = \tilde{A}^{R}_{d,1}$ and $\tilde{A}^{I}_{2} = \tilde{A}^{I}_{1}$, and therefore its clear that once again no second unitary transformation is necessary, and that

\begin{equation}\label{conjugationtwo}
\vec{a}_{2}\vec{a}^{\dagger}_2 = G_{(U^{(2)})}\big[A^{(2)}(\theta_2,\vec{\alpha}_2^R,\vec{\alpha}_2^{I})\big]G_{(U^{(2)})}^T,
\end{equation}
where $\vec{\alpha}_2^R = \vec{\alpha}_1^R$, $\vec{\alpha}_2^I = \vec{\alpha}_1^I$ and we have defined $U^{(2)} = U^{(2)\dagger}_1$ and $G_{(U^{(2)})} = \hat{\mathrm{Ad}}(U^{(2)})$. Finally, for any channel $T_2(t) = \mathrm{exp}(t\mathcal{L}_{\vec{a}_2\vec{a}_2^{\dagger}})$ from the semigroup generated by $\vec{a}_2\vec{a}_2^{\dagger}$,

\begin{equation}\label{chan2}
T_2(t)(\rho) =    U^{(2)\dagger}\bigg(T_{A^{(2)}}(t)\big[U^{(2)}\rho U^{(2)\dagger}\big]\bigg)U^{(2)},
\end{equation}
where $T_{A^{(2)}}(t) = \mathrm{exp}(t\mathcal{L}_{A^{(2)}})$. 

At this stage, given $\epsilon >0$, $t>0$ and $\rho(0)$, in order to efficiently implement $T(t) = e^{t\mathcal{L}}$ one constructs $S_{2k}(\hat{\mathcal{L}}_1,\ldots,\hat{\mathcal{L}}_m,t/r)$ as per \eqref{suz2}, with $k$ given by \eqref{bound1}. One then implements $S_{2k}$ $rL_1$ times, with $r$ given by \eqref{bound2}, and each call to $T_k(\tilde{t})$ is achieved using the unitary conjugation of some channel from the universal set, as per \eqref{chan1} and \eqref{chan2}, where $\tilde{t}$ incorporates $\gamma_k$.

\section{Conclusion}\label{conclusion}

Utilising the composition framework of linear combination and unitary conjugation we have constructed a universal set of generators for the simulation of Markovian semigroup dynamics. More precisely, we have constructed a $d^2-3$ parameter family of semigroup generators, such that any Markovian semigroup describing the dynamics of a $d$ dimensional Markovian open quantum system can be simulated through the implementation of unitary operations on the system and quantum channels from the semigroups generated by the $d^2-3$ parameter family of generators. Furthermore, assuming the ability to implement all operations from the universal set, the construction of such a universal set implies a natural methodology for the simulation of Markovian open quantum systems: Given such a system, one utilises the construction of Theorem \ref{main} to decompose the generator of the system into the linear combination and unitary conjugation of generators from the universal set, before utilising Suzuki-Lie-Trotter techniques \cite{Sweke2014, Papageorgiou2012} to simulate the original system through simulations of the constituent semigroups. This approach will provide a method for the \textit{efficient} simulation, with respect to this universal set, of any Markovian open quantum system for which the number of distinct physical dissipation processes with non-zero rates (the number of Lindblad operators) scales polynomially with the number of subsystems.

Given such a methodology, it is clear that in order to use this approach for the simulation of  arbitrary Markovian open quantum systems one need only to focus on explicitly constructing methods, and developing the experimental capability, for efficiently simulating those systems whose generators are specified by GKS matrices belonging to the universal set of Theorem \ref{main}. These generators provide a significant simplification from the general case, and in principle, these systems could be simulated using any of the previous methods \cite{Barthel2012, Tempel2014, DiCandia2014, Muller2011, Schindler2013, Barreiro2011, Weimer2011, Weimer2010} for the simulation of Markovian open quantum systems. Another appealing approach would be to investigate the possibility of utilising the inherent dissipation and decoherence within currently available controllable quantum devices for the implementation of non-unitary processes from the universal set, therefore developing the potential of quantum simulators other than universal quantum computers. One other possibility, already explored in detail for the single qubit case \cite{Sweke2014}, would be to explicitly construct parametrised descriptions of the quantum channels appearing in the semigroup generated by an arbitrary element of the universal set. Given such an explicit parametrised family of quantum channels, the methods of \cite{Wang2014} could be used to implement any such channel for any given time, on a minimal dilation space, through the simulation of constituent extreme channels.

Given these results, a natural open question concerns the extension of this approach for more general open quantum systems, such as those described by time-dependent generators \cite{Kliesch2011}. In order to extend this approach one could investigate the possibility of utilising more general composition frameworks which are not constrained to preserve Markovianity, possibly including feedback \cite{Lloyd2001} or probabilistic implementations of quantum channels \cite{Wang2013, Wang2014}.

 \begin{acknowledgments}
This work is based upon research supported by the South African Research Chair Initiative of the Department of Science and  Technology and National Research Foundation. RS acknowledges the financial support of the National Research Foundation SARChI program. The work by DB was supported in part by the ANR contract ANR-14-CE25-0003. The authors would like to thank Tongyang Li and Andrew Childs for bringing to our attention problems with generalisations of higher-order Suzuki integrators into the superoperator setting.
\end{acknowledgments}

\begin{appendix}
\section{Simulation of linear combinations}\label{lincomb}

Given the generator of a Markovian semigroup, $\mathcal{L} = \sum_{j=1}^m\mathcal{L}_j$, we want to show that for any $t>0$ it is possible to implement $T(t) = e^{t\mathcal{L}}$, up to arbitrary accuracy $\epsilon >0$, using only $\mathrm{poly}\big(||\mathcal{L}||_{(1\rightarrow 1)},t,1/\epsilon, m \big)$ number of implementations of quantum channels $T^{(j)}(t') = e^{t'\mathcal{L}_j}$. Using Suzuki-Lie-Trotter techniques \cite{Trotter1959, Suzuki1990, Suzuki1991} the analogous problem for linear combinations of Hamiltonians has been studied extensively \cite{Sanders2013, Berry2015, Papageorgiou2012, Wiebe2010},  and generalisations to the context of open quantum systems have been considered before for both the case of time-dependent \cite{Kliesch2011} and time-independent \cite{Sweke2014} generators. Here we present a direct generalisation of the work in \cite{Papageorgiou2012} to the super-operator setting, first presented in \cite{Sweke2014}, which provides the best current bounds on the number of implementations of quantum channels $T^{(j)}(t') = e^{t'\mathcal{L}_j}$ required, within the context of time-independent generators $\mathcal{L}$.

We begin by assuming that 
\begin{equation}
||\mathcal{L}_1||_{(1\rightarrow 1)} \geq ||\mathcal{L}_2||_{(1\rightarrow 1)} \geq \cdots \geq ||\mathcal{L}_m||_{(1\rightarrow 1)}
\end{equation}
and defining the normalised component generators $\hat{\mathcal{L}}_j = \mathcal{L}_j/L_1$, where we have defined $L_j := ||\mathcal{L}_j||_{(1\rightarrow 1)} $ for all $j$. We then follow \cite{Papageorgiou2012} and define the basic Lie-Trotter product formula \cite{Trotter1959} as,

\begin{align}
S_{2}(\hat{\mathcal{L}}_1,\ldots,\hat{\mathcal{L}}_m,\lambda) &= \prod_{j = 1}^m e^{(\frac{\lambda}{2})\hat{\mathcal{L}}_{j}}\prod_{j' = m}^1 e^{(\frac{\lambda}{2})\hat{\mathcal{L}}_{j'}}\\ \label{suz1}
&=\prod_{j = 1}^m T^{(j)}(t_\lambda)\prod_{j' = m}^1 T^{(j')}(t_\lambda),
\end{align}
where $t_\lambda = \lambda/(2L_1)$. Suzuki's higher order integrators \cite{Suzuki1990,Suzuki1991}  are then defined using the recursion relation

\begin{equation}\label{suz2}
S_{2k}(\lambda) = [S_{2k-2}(p_k\lambda)]^2[S_{2k-2}((1 - 4p_k)\lambda)] [S_{2k-2}(p_k\lambda)]^2,
\end{equation}
where $p_k = (4 - 4^{1/(2k-1)})^{-1}$ for $k >1$ and for notational convenience we have used $S_{2k}(\lambda)$ and $S_{2k-2}(\lambda)$ to denote $S_{2k}(\hat{\mathcal{L}}_1,\ldots,\hat{\mathcal{L}}_m,\lambda)$ and $S_{2k-2}(\hat{\mathcal{L}}_1,\ldots,\hat{\mathcal{L}}_m,\lambda)$ respectively. At this stage it is essential to note that for $k > 1$ we have $(1 - 4p_k) < 0$, and therefore applying the recursion rule \eqref{suz2} allows us to see that for all $k >1$ implementation of $S_{2k}(\lambda)$ requires the simulation of multiple propagators $T^{(j)}(\tilde{t})$ with $\tilde{t} < 0$ \cite{Suzuki1990}. As such propagators are \emph{not} quantum channels (in particular they may violate complete positivity, or even positivity) \cite{Petruccione2002}, we therefore restrict ourselves here to first order ($k = 1$) integrators. This is in juxtaposition to the Hamiltonian simulation case, where for generators $\mathcal{L}_j(\cdot) = -i[H_j,\cdot]$ of purely coherent evolution, the propagators $T^{(j)}(\tilde{t}) = e^{\tilde{t}\mathcal{L}_j}$ are quantum channels (in fact unitary conjugations) even for the case of $\tilde{t} <0$.

In light of these considerations, we therefore proceed to examine the efficiency of approximating $T(t) = \mathrm{exp}(t\sum_{j = 1}^m\mathcal{L}_j)$ with sequences of quantum channels of the form $[S_2(t/r)]^x$. In particular, we note that $S_2(\lambda)$ consists of the product of $2m-1$ exponentials, and hence we can define

\begin{equation}\label{nexp}
N_{\mathrm{exp}} = (2m-1)x
\end{equation}
as the number of exponentials, and hence quantum channels, in the expression $[S_2(t/r)]^x$. The following theorem \cite{Sweke2014}, a direct generalization of the work in \cite{Papageorgiou2012} to the superoperator setting, then gives the desired result

\begin{theorem}\label{baset}
Let $1 \geq \epsilon > 0$ be such that $(9/2)L_2mt \geq \epsilon$, then for 

\begin{equation}
r \geq \frac{\sqrt{2L_2}(mt)^{3/2}}{\epsilon^{1/2}},
\end{equation}
we have that

\begin{equation}\label{bound}
\Big|\Big|\mathrm{exp}\Big( t\sum_{j = 1}^m\mathcal{L}_j\Big) - \big[S_{2}(\hat{\mathcal{L}}_1,\ldots,\hat{\mathcal{L}}_m,t/r) \big]^{rL_1}\Big|\Big|_{1\rightarrow 1}\leq \epsilon,
\end{equation}
and the number of exponentials required is bounded by

\begin{equation}\label{res}
\mathrm{N_{exp}} \leq (2m-1)  \frac{\sqrt{2L_2}L_1(mt)^{3/2}}{\epsilon^{1/2}}.
\end{equation}
\end{theorem}

Furthermore, by definition of the $(1\rightarrow 1)$ norm we have that for any density matrix $\rho$ and any superoperators $P$ and $Q$,

\begin{equation}
||P(\rho) - Q(\rho)||_1  \leq ||P-Q||_{1\rightarrow 1}
\end{equation}
and as such the results of Theorem \ref{baset} bound the error in the output state obtained when approximating $T(t)$ with $[S_{2}(t/r)]^{rL_1}$. 

\section{Properties of the adjoint representation}\label{adjointap}

We summarise here properties and characterisations of the adjoint representation of $\mathrm{SU}(d)$, in order to set the notation used for a rigorous description of the effect of unitary conjugation and to provide the fundamental results used in the proof of our main result. For more detail, and proofs of the statements which follow, the interested reader is referred to \cite{Fulton1991, Lee2013}.

For any Lie group $G$, with Lie algebra $\mathfrak{g}$, we define the conjugation map 

\begin{equation}
\psi_g:G \rightarrow G
\end{equation}
via
\begin{equation}
\psi_g(h) = ghg^{-1}
\end{equation}
$\forall g,h \in G$. The adjoint representation of G,

\begin{equation}
\mathrm{Ad}: G \rightarrow \mathrm{GL}(\mathfrak{g}),
\end{equation}
is then defined via
\begin{equation}
\mathrm{Ad} (g) := \mathrm{d}\psi_g\big|_e :\mathfrak{g}\rightarrow\mathfrak{g},
\end{equation}
where $\mathrm{d}\psi_g\big|_e$ is the differential of $\psi_g$ at the identity element of $G$. The adjoint representation of $\mathfrak{g}$,

\begin{equation}
\mathrm{ad}: \mathfrak{g} \rightarrow \mathrm{End}(\mathfrak{g}) \simeq \mathfrak{gl}(\mathfrak{g}),
\end{equation}
is then induced from $\mathrm{Ad}$ and defined via,
\begin{equation}
\mathrm{ad}(X) = \mathrm{d}(\mathrm{Ad})\big|_e(X) : \mathfrak{g}\rightarrow\mathfrak{g}
\end{equation}
$\forall X \in \mathfrak{g}$. We then define $\mathrm{Int}(\mathfrak{g}) = \mathrm{Im}(\mathrm{Ad}) \subseteq \mathrm{GL}(\mathfrak{g})$, the image of $\mathrm{Ad}$, and $\mathfrak{Int}(\mathfrak{g}) = \mathrm{Im}(\mathrm{ad}) \subseteq \mathfrak{gl}(\mathfrak{g})$, the image of $\mathrm{ad}$. One can show that $\mathrm{Ad}$ is a Lie group homomorphism, $\mathrm{ad}$ a Lie algebra homomorphism, and that $\mathrm{Int}(\mathfrak{g})$ is a Lie group with Lie algebra $\mathfrak{Int}(\mathfrak{g})$.

As we will be concerned with $\mathrm{SU}(d)$, we assume here that $\mathfrak{g} \simeq \mathbb{R}^n$ is a real vector space (where for $\mathfrak{g} = \mathrm{su}(d)$ we have that $n = d^2-1$). Under this assumption, let $\{X_i\}|_{i=1}^n$ be a basis for $\mathfrak{g}$, with structure constants

\begin{equation}
[X_i,X_j] = f_{ijk} X_k,
\end{equation}
where we have utilised the summation notation for repeated indices. Furthermore, for arbitrary $X \in \mathfrak{g}$, let $X = x_i X_i$, so that by identifying basis elements we can define the natural linear isomorphism

\begin{equation}
f: \mathfrak{g} \rightarrow \mathbb{R}^n,
\end{equation}
such that $f(X) = \vec{x}$. Given this, it is possible to show that $\forall X,Y \in \mathfrak{g}$,

\begin{equation}
\mathrm{ad}(X)(Y) = [X,Y],
\end{equation}
such that via linearity of the Lie bracket

\begin{equation}
\mathrm{ad}(X)(Y) = \big[x_i \mathrm{ad}(X_i)\big](Y).
\end{equation}
Furthermore, if $\mathrm{ker}(\mathrm{ad}) = 0$, which is indeed the case for $\mathfrak{g} = \mathrm{su}(d)$, then $\mathrm{ad}: \mathfrak{g} \rightarrow \mathfrak{Int}(\mathfrak{g})$ is also a linear isomorphism, such that $\{\mathrm{ad}(X_i)\}|_{i = 1}^n$ is a basis for the Lie algebra $\mathfrak{Int}(\mathfrak{g})$.
Using the structure constants, for any $X \in \mathfrak{g}$ we can then define $\hat{\mathrm{ad}}(X) \in \mathcal{M}_n(\mathbb{C})$, the matrix representation of $\mathrm{ad}(X)$, such that $\forall Y \in \mathfrak{g}$

\begin{equation}
\mathrm{ad}(X)(Y) =f^{-1}\big(\hat{\mathrm{ad}}(X)f(Y)\big),
\end{equation}
via $\hat{\mathrm{ad}}(X) = x_i \hat{\mathrm{ad}}(X_i)$, where the matrix elements of $\hat{\mathrm{ad}}(X_i)$ are given by 
\begin{equation}
[\hat{\mathrm{ad}}(X_i)]_{jk} = f_{ijk}.
\end{equation}
In addition, one can show that $\forall X \in \mathfrak{g}$ the $\mathrm{Ad}$ map satisfies

\begin{equation}
\mathrm{Ad}\big(\mathrm{exp}(X)\big) = \mathrm{exp}\big({\mathrm{ad}(X)}\big),
\end{equation}
and that for connected matrix groups $G$ (such as $\mathrm{SU}(d)$)
\begin{equation}
\mathrm{Ad}(g)(Y) = gYg^{-1},
\end{equation}
$\forall g \in G$ and $\forall Y \in \mathfrak{g}$, such that for any $g = \mathrm{exp}(X) \in G$ we have the equivalence

\begin{align}
\mathrm{Ad}(g)(Y) &= f^{-1}\big(e^{\hat{\mathrm{ad}}(X)}f(Y) \big)\nonumber\\ &= f^{-1}\big(\hat{\mathrm{Ad}}(g)f(Y) \big)\nonumber\\\label{equivalence} &=  gYg^{-1},
\end{align}
where $\hat{\mathrm{Ad}}(g) \in \mathcal{M}_{n}(\mathbb{C})$ is the matrix representation of $\mathrm{Ad}(g)$.

\end{appendix}


\end{document}